\documentclass[preprint]{ptephy_v1}
\usepackage{bm,latexsym,amssymb,amsmath,mathrsfs}
\usepackage{color}
\usepackage{times,setspace}
\usepackage{ulem}

\preprintnumber{RIKEN-iTHEMS-Report-22}

\begin{document}

\title{Toward nuclear physics from lattice QCD on quantum computers}
\author{Arata Yamamoto$^{a}$}
\author{Takumi Doi$^{b}$}
\affil{$^{a}$Department of Physics, The University of Tokyo, Tokyo 113-0033, Japan}
\affil{$^{b}$Interdisciplinary Theoretical and Mathematical Sciences Program (iTHEMS), RIKEN, Wako 351-0198, Japan}

\begin{abstract}
One of the ultimate missions of lattice QCD is to simulate atomic nuclei from the first principle of the strong interaction.
This is an extremely hard task for the current computational technology, but might be reachable in coming quantum computing era.
In this paper, we discuss the computational complexities of classical and quantum simulations of lattice QCD.
It is shown that the quantum simulation scales better as a function of
a nucleon number and thus will outperform for large nuclei.
\end{abstract}

\subjectindex{D34}

\maketitle

\section{Introduction}

Quantum computing era is coming around the corner.
When quantum computation is put to practical use, paradigm shifts would take place in computational studies of hadron and nuclear physics.
One of the most expected fields is lattice quantum chromodynamics (QCD).
Lattice QCD researchers have an ambition to study
various important physics which suffer from the sign problem in classical computations,
e.g., real-time dynamics \cite{Martinez:2016yna,Klco:2018kyo,Klco:2019evd,Gustafson:2019vsd,Kharzeev:2020kgc,Mathis:2020fuo,Yamamoto:2020eqi,Hayata:2021kcp,ARahman:2021ktn,Zhou:2021kdl,Mildenberger:2022jqr,Atas:2022dqm,Farrell:2022vyh}
and properties of dense quark matters \cite{Yamamoto:2021vxp,Tomiya:2022chr,Czajka:2022plx}.
The applicability of quantum computations is not limited to the sign problem,
and it is important to open a new avenue to study unsolved problems in lattice QCD.

A nuclear many-body system is one of such challenging problems in lattice QCD.
The nuclear many-body system can be, in principle, simulated by lattice QCD because it is just a composite state of quarks and gluons.
The simulation is however computationally difficult even with state-of-the-art supercomputers.
As will be discussed later, the computational cost for the corresponding
multi-nucleon correlation function is known to blow up quickly as a nucleon number increases.
In addition, there exists a so-called signal-to-noise problem,
where a correlation function suffers from
an exponentially larger statistical error with an increased nucleon number~\cite{Lepage:1989hd},
and this issue is related to the sign problem in finite density.
Thanks to the significant progress over the decades,
the current front-line reaches the simulations of a few-body systems:
For instance,
two-baryon interactions near the physical point~\cite{Gongyo:2017fjb, HALQCD:2018qyu, HALQCD:2019wsz, Lyu:2021qsh, Aoki:2020bew}
and three-nucleon interactions at heavy quark masses~\cite{Doi:2011gq, Aoki:2020bew}
are obtained from the HAL QCD method~\cite{Ishii:2006ec, Ishii:2012ssm}.
See also recent calculations of two-baryon interactions
at heavy quark masses~\cite{Francis:2018qch, Horz:2020zvv, Amarasinghe:2021lqa}
from the finite volume method~\cite{Luscher:1990ux}.
It is, however, still far out of reach of current computations
to go beyond few-body systems to determine the properties of large nuclei or many-body forces
directly from QCD.

In this paper, we discuss nuclear many-body problem in lattice QCD on quantum computers.
The most important question is whether the quantum computation overcomes the conventional one on classical computers.
To answer this question, we estimate their computational complexities in Sec.~\ref{secC}.
We also show some results of small-scale noiseless quantum simulation in Sec.~\ref{secD}.
Since the Hilbert space of QCD is too large, we employ a toy model of nucleons.
Finally, in Sec.~\ref{secF}, we conclude the paper with some future applications.

\section{Computational cost comparison}
\label{secC}

We compare the computational cost of classical and quantum simulations of lattice QCD.
To make the argument concrete, let us consider the simulations for the ground state mass of one nucleus.
We estimate how the simulation cost scales as a function of the three-dimensional spatial volume $V$ and the nucleon number $A$.
We treat them as independent parameters, although they are not independent in the sense that the spatial volume must be taken larger than the physical size of the nucleus.
We consider quark masses at the physical point and estimate the cost with the lattice spacing $a$ fixed.
We implicitly assume the Wilson fermion formalism although it is not essential for the scaling argument below.
The Wilson fermion has an advantage in nuclear physics because flavor number is tunable.
The general advantages of the Wilson fermion on quantum computing are described in the literature \cite{Zache:2018jbt}.

\subsection{Classical simulation}

The conventional lattice QCD calculation is based on the path integral formalism.
The lattice QCD action is given by
\begin{equation}
 S = S_{\rm quark}[\bar{\psi},\psi,U] + S_{\rm gluon}[U] ,
\end{equation}
where $\psi$ and $\bar{\psi}$ are Grassmann quark fields and $U$ is a gauge link field.
The quark fields have internal spinor, color, and flavor spaces and the link field has an internal color space and four directions, but the indices are omitted for notational simplicity.
The path integral is given by 
\begin{equation}
 Z = \int D\bar{\psi} D\psi DU \ e^{-S}
\end{equation}
and the expectation value of the observable ${\cal O}$ is calculated by 
\begin{equation}
  \langle {\cal O} \rangle =   \frac{1}{Z} \int D\bar{\psi} D\psi DU \ {\cal O} \ e^{-S}  .
\end{equation}
They are defined on a four-dimensional lattice $V\times N_\tau$.

For the calculation of a nucleus with a mass number $A$, one typically considers a two-point correlation function of nucleus composite fields.
The nucleus composite field is defined from $A$ baryon fields as
\begin{equation}
  \mathcal{N}(\tau) = C_\mathcal{N} \mathcal{B}(x_1,\tau) \cdots \mathcal{B}(x_A,\tau).
  \label{eq:nuc_op}
\end{equation}
The coefficient $C_\mathcal{N}$ is a matrix of spin and flavor such that $\mathcal{N}$ represents the nucleus of interest.
The baryon field is defined from three quark fields as
\begin{equation}
  \mathcal{B}(x,\tau) = C_\mathcal{B} \psi(x,\tau) \psi(x,\tau) \psi(x,\tau) .
  \label{eq:quark_op}
\end{equation}
The coefficient $C_\mathcal{B}$ is a matrix of spinor, color, and isospin such that $\mathcal{B}$ represents a color-singlet proton or neutron with spin $1/2$.
The ground state mass $M$ can be obtained from the temporal correlation
\begin{equation}
 \langle {\mathcal{N}}(\tau) \bar{\mathcal{N}}(0) \rangle \propto e^{-M\tau}.
\end{equation}
In actual calculations, one usually repeats the calculation of correlation functions
with various choices for $x_1, \cdots, x_A$
(or similar choices in the momentum space)
with zero-momentum projection for the total system.
As a simplest choice,
we consider the spatial summation for the center of mass coordinate of the sink field in the correlation function,
and count the corresponding computational cost.
Spatial summation usually leads to the implicit improvement of statistical fluctuations with larger $V$,
but we do not consider such effect in the estimate.

The computational cost is estimated as follows.
\begin{description}
\item[{\bf Quark propagator}]\
The quark propagators are computed by conjugate gradient (CG) based algorithms.
The convergence of the CG based algorithms is generally dependent on a condition number but asymptotically independent of $V$.
This is because the condition number is given by the difference between the maximum and minimum eigenvalues and they are bounded by lattice cutoff and quark mass, respectively.
When the convergence is independent of $V$, the computational cost for one propagator is proportional to the matrix size $V$.

\item[{\bf Gauge configuration}]\
The path integral is numerically evaluated by the Hybrid Monte Carlo (HMC) method.
As the volume becomes larger, the time step of the evolution of molecular dynamics (MD)
should be taken finer, which increases the computational cost with a scaling of
$O(V^\frac14)$ in the case of a second-order MD integrator as leap-frog or Omelyan~\cite{Montvay1997}.
The scaling can be slightly improved with higher-order MD integrator, but
the difference is marginal in the following discussions.
Since the computational cost of each time step scales as $O(V)$ due to the calculation
of propagators,
the total cost for generating one gauge configuration is
$t_1 = O(V^\frac54)$~\cite{Montvay1997, Luscher:2010ae}.

\item[{\bf Correlation function}]\
The computation of a correlation function involves
that of quark propagators and
that of contractions.
The cost for the propagators
with all $x_1, \cdots, x_A$ is $t_2 = O(V A)$.
The contractions consist of 
two kinds of calculations:
Wick contraction and color/spinor contraction.
The former corresponds to quark permutation and
thus the cost of the straightforward calculation scales
factorially with $A$, $t_{3a} \simeq O((\frac{3}{2}A!)^2)$.
On the other hand, the latter corresponds to the contraction
for internal degrees of freedom, whose cost scales
$t_{3b} = O(e^{cA})$~\cite{Doi:2012xd},
where $c$ is a constant which depends on the details of the computational setup and algorithm.
We also consider the cost for the zero-momentum projection, $t_{3c} = O(V)$.
Therefore, the cost of straightforward contraction calculation amounts to be
$t_3 = t_{3a} \times t_{3b} \times t_{3c} \simeq O(V\cdot(\frac{3}{2}A!)^2 \cdot e^{cA})$.
In the last decade, significant algorithmic development to reduce the
cost of contraction have been achieved~\cite{Yamazaki:2009ua, Doi:2012xd, Detmold:2012eu, Gunther:2013xj, Nemura:2015yha, Horz:2019rrn},
which are crucial to perform lattice QCD studies for few-baryon systems.
If one further considers the scaling with a much larger $A$,
the most significant cost is the factorial factor associated with the permutation,
and the algorithm which can reduce this cost to $t_{3a} = O(A^3)$~\cite{Kaplan:2007DWF}
would be most efficient.
Therefore, we take the cost of the contraction as $t_3 = O(V A^3 e^{cA})$.

\item[{\bf Statistical error}]\
It is known that statistical error becomes exponentially worse
in a larger $A$ system~\cite{Lepage:1989hd}.
Therefore, it is necessary to increase the number of configurations
to keep a relative statistical error $O(1)$
by the scaling of $N_{\rm conf} = O(e^{c'A})$,
where 
$c'$ is a number which linearly depends on
the Euclidean time $\tau$ in the correlation function.
We take $\tau$ to be independent of $V$ and $A$, since the energy gap
between the ground state and the first excited state
is rather insensitive to $V$ and $A$,
so $c'$ becomes constant.
\end{description}
Combining all the estimates above, the total cost scales as
\begin{equation}
\label{eqcs}
(t_1+t_2+t_3)\times N_{\rm conf} = O(V^\frac54 e^{c_1 A}) + O(V e^{c_2 A} A^3)
\end{equation}
in the classical simulation,
where $0 < c_1 (\equiv c') < c_2 (\equiv c+c')$ are some constants
which depend on details of computations.

\subsection{Quantum simulation}
\label{secqc}

The Hamiltonian formalism is favored for quantum simulation.
The lattice QCD Hamiltonian is the sum of the quark and gluon parts,
\begin{equation}
 \hat{H}  = \hat{H}_{\rm quark}[\hat{\psi}^\dagger,\hat{\psi},\hat{U}] + \hat{H}_{\rm gluon}[\hat{E},\hat{U}].
\end{equation}
The quark Hamiltonian is given by the quark creation and annihilation operators, $\hat{\psi}^\dagger,\hat{\psi}$, and the spatial link operator, $\hat{U}$.
The gluon Hamiltonian is given by the chromoelectric operator, $\hat{E}$, and the spatial link operator, $\hat{U}$.
These operators can be constructed by sequential manipulations of fundamental gates.
The gate costs are $O(1)$ for the gluon operators and $O(\log V)$ for the quark operators \cite{2002AnPhy.298..210B},
but we neglect the logarithmic correction on the following estimation.
The Hilbert space is the direct product of the quark and gluon parts. 
The quark Hilbert space is $2^{D_{\rm quark}}$ dimensional with $D_{\rm quark}=V$(site)$\times 4$(spinor)$\times 3$(color)$\times 2$(flavor) and the gluon Hilbert space is $d^{D_{\rm gluon}}$ dimensional with $D_{\rm gluon}=3V$(link).
The number of gluon modes $d$ is infinite in principle, but truncated by a finite number in numerical simulation.
Although the Hilbert space can be reduced by gauge fixing or solving the Gauss law, we do not consider such a reduction here.

We adopt the quantum adiabatic algorithm \cite{RevModPhys.90.015002}. 
The algorithm consists of three steps: state preparation, adiabatic evolution, and measurement.
First, we prepare the ground state of a solvable Hamiltonian $\hat{H}_0$ for given proton and neutron numbers.
The choice for $\hat{H}_0$ is not unique.
One example is 
\begin{equation} 
\label{eqH0}
\hat{H}_0 = \sum_x \hat{\psi}^\dagger(x) \left\{ m\gamma^0 + v(x)\right\} \hat{\psi}(x) - w \sum_{\rm plaq} {\rm ReTr } \hat{U}\hat{U}\hat{U}\hat{U},
\end{equation}
where $m$ is the quark mass, $w$ is a positive constant, and $\sum_{\rm plaq}$ is the summation over plaquettes.
The potential term $v(x)$ is introduced to resolve the ground state degeneracy and non-existence of a quantum phase transition is assumed between the ground states of $\hat{H}_0$ and $\hat{H}$ \footnote{For the validity of the quantum adiabatic algorithm, the ground state must be gapped during the evolution \eqref{eqQAC}. There must be the one-to-one correspondence between the ground states of $\hat{H}_0$ and $\hat{H}$, otherwise the gap closes somewhere and the algorithm breaks down. Also, if the evolution passes through a quantum phase transition, the gap goes to exponentially small and the convergence is exponentially slow. Fortunately, lattice QCD has no quantum phase transition between weak coupling limit and strong coupling limit.}.
We define the nucleus operator $\hat{\mathcal{N}}$ from $A$ baryon operators,
whereas each baryon operator $\hat{\mathcal{B}}(x)$ is defined from three quark operators,
in the same way as Eqs.~(\ref{eq:nuc_op}) and (\ref{eq:quark_op}) except that the argument $\tau$ does not exist.
The nucleus operator $\hat{\mathcal{N}}$ is constructed such that the ground state of $\hat{H}_0$ is given by
\begin{equation}
 |\Phi_A\rangle = \hat{\mathcal{N}}^\dagger |\Phi_0\rangle,
\end{equation}
where $|\Phi_0\rangle$ is the ground state of $\hat{H}_0$ for $A=0$, i.e., the vacuum.
The ground state of the Hamiltonian $\hat{H}$ is obtained by the evolution equation
\begin{equation}
 |\Psi_A\rangle = \exp\left(-i \int_0^S ds \hat{h}(s) \right) |\Phi_A\rangle ,
\label{eqQAC}
\end{equation}
where $\hat{h}(s)$ interpolates $\hat{H}_0$ at $s=0$ and $\hat{H}$ at $s=S$.
The simplest choice is
\begin{equation}
 \hat{h}(s) = \left( 1-\frac{s}{S} \right) \hat{H}_0 + \frac{s}{S} \hat{H}
\end{equation}
although the interpolation function is not unique in general.
As long as $\hat{H}_0$ conserves up and down quark numbers, the evolution equation \eqref{eqQAC} does not change proton and neutron numbers.
We can obtain the ground state $|\Psi_A\rangle$ with the same proton and neutron numbers as the initial state $|\Phi_A\rangle$.
The ground state mass is defined by the energy difference from the vacuum,
\begin{equation}
 M = \langle \Psi_A| \hat{H} |\Psi_A\rangle - \langle \Psi_0| \hat{H} |\Psi_0\rangle .
\end{equation}

The computational cost is estimated as follows.
\begin{description}
\item[{\bf State preparation}]\
For the initial Hamiltonian \eqref{eqH0}, the vacuum $|\Phi_0\rangle$ is constructed by making all the spatial links be unity and all the antiquarks occupy over the whole lattice.
The manipulation of $\hat{\mathcal{N}}^\dagger$ is given by making $A$ baryons occupy the sites from lower energy of $v(x)$.
The sum of the cost is $T_1 = O(V)+O(A)$.

\item[{\bf Adiabatic evolution}]\
For the validity of the adiabatic evolution \eqref{eqQAC}, $S$ is large enough that
\begin{equation}
\label{eqSmax}
 S \gg \max_{s\in [0,S]} \frac{ || \hat{H}-\hat{H}_0 || }{ \{ \Delta(s) \}^{2} }.
\end{equation}
The lower bound is sensitive to the smallest value of the spectral gap $\Delta(s)$ between the ground state and the first excited state of $\hat{h}(s)$.
The gap around $s=0$ can be controlled and taken arbitrarily large by the choice of $\hat{H}_0$, while the gap around $s=S$ is given by the physical energy gap of $\hat{H}$.
If $\hat{h}(s)$ can be optimized, the physical energy gap will be a bottleneck.
Two kinds of physical excitation are possible: the internal excitation and the central motion of a nucleus.
The internal excitation is independent of $V$ and its energy $\Delta_{\rm int}$ is typically a few MeV.
The central motion depends on $V$ and the minimum kinetic energy is given by
\begin{equation}
\label{eqdelta}
 \Delta_{\rm cen} = \frac{1-\cos\left(\frac{2\pi}{L}\right)}{Ma^2} = O(V^{-\frac23}A^{-1})
\end{equation}
in an isotropic cube $V=L^3$.
An example value is $\Delta_{\rm cen}\simeq 0.4$ MeV for $V=128^3$, $a^{-1}=2$ GeV, and $A=12$.
When $V$ and $A$ are large, $\Delta_{\rm cen}$ is smaller than $\Delta_{\rm int}$.
Thus $\Delta_{\rm cen}$ gives the upper bound of asymptotic scaling.
The matrix norm $||\hat{H}-\hat{H}_0||$ of Eq.~\eqref{eqSmax} scales as $O(V)$.
Therefore the lower bound of $S$ is given by $O(V\Delta^{-2}_{\rm cen})$.
Since the circuit depth of the evolution \eqref{eqQAC} is proportional to $S$ and one manipulation of $\hat{h}(s)$ is $O(V)$, the cost is estimated as
\begin{equation}
\label{eqtii}
 T_2 = O(V^2\Delta^{-2}_{\rm cen}) = O(V^{\frac{10}3}A^2).
\end{equation}
If the central motion excitation is absent, the convergence is bounded by $\Delta_{\rm int}$ and the scaling is improved as $T'_2 = O(V^2\Delta^{-2}_{\rm int}) = O(V^2)$.
When $\hat{H}_0$ is chosen to be translationally invariant like the free Dirac Hamiltonian
\footnote{There are however two concerns of the free Dirac Hamiltonian.
One is degeneracy.
The ground state of a finite number of free quarks is in general non-unique.
The other is the nuclear liquid-gas phase transition.
Nuclei are liquid drops while free quarks are gas.
The evolution might across the liquid-gas phase transition.},
the evolution equation \eqref{eqQAC} conserves the total momentum.
We can exclude the central motion excitation by setting the initial total momentum to zero.
Since the general construction of translationally-invariant $\hat{H}_0$, other than the free Dirac Hamiltonian, is not known, further studies on this point are desirable in future.

\item[{\bf Measurement}]\
 The energy is measured as the expectation value of the Hamiltonian over the volume, so the cost is $T_3 = O(V)$.

\begin{figure*}[t]
\begin{minipage}{0.45\textwidth}
\begin{center}
 \includegraphics[width=1\textwidth]{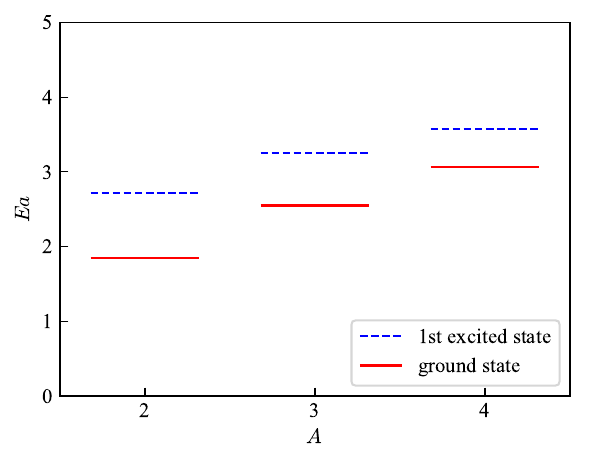}
\end{center}
\end{minipage}
\begin{minipage}{0.45\textwidth}
\begin{center}
 \includegraphics[width=1\textwidth]{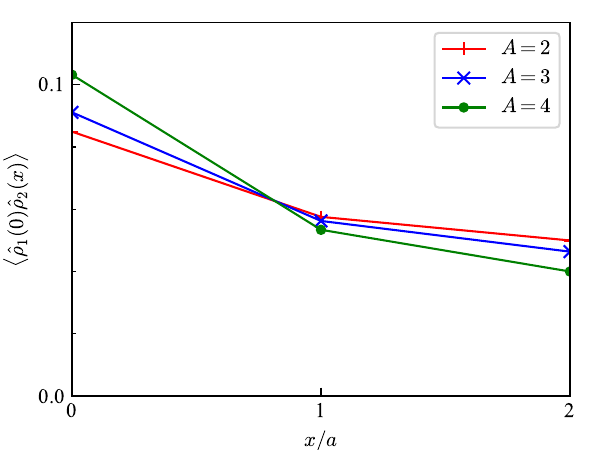}
\end{center}
\end{minipage}
\caption{
\label{figE}
Exact diagonalization results of the energy eigenvalues $E$ (left) and the ground-state correlation function $\langle \hat{\rho}_1(0) \hat{\rho}_2(x)\rangle$ (right).
The parameters are $ma=1$ and $\lambda a=0.5$ and the lattice volume is $V=L=4$.
}
\end{figure*}

\item[{\bf Quantum fluctuation}]\
One circuit is a sequential execution of the above steps and the circuit is executed $N_{\rm shot}$ times to take the average $E=\langle \hat{H} \rangle$.
The statistical average is accompanied by the statistical error
$\delta E = \sqrt{(\langle \hat{H}^2 \rangle - \langle \hat{H} \rangle^2)/(N_{\rm shot}-1)}$.
If the theory is located at a quantum critical point, quantum fluctuation diverges, so $\delta E$ diverges.
The circuit must be executed exponentially many times to obtain an order-one relative statistical error, $\delta E/E=O(1)$, as the volume increases.
In nuclear many-body problems, however, we do not encounter a quantum critical point (except for the liquid-gas critical point of nuclear matter).
The relative statistical error is insensitive to $V$ and $A$. 
Thus $N_{\rm shot} = O(1)$.
\end{description}
The total cost is given by
\begin{equation}
\label{eqqs}
 (T_1+T_2+T_3) \times N_{\rm shot} = O(V^{\frac{10}3}A^2).
\end{equation}
in the quantum adiabatic algorithm.
The quantum scaling \eqref{eqqs} is at least exponentially better than the classical scaling \eqref{eqcs} as a function of $A$.
If the central motion excitation can be excluded, Eq.~\eqref{eqqs} will be improved to $(T_1+T'_2+T_3) \times N_{\rm shot} = O(A)+O(V^2)$.

\section{Model demonstration}
\label{secD}

Unfortunately, we cannot numerically check the asymptotic scaling \eqref{eqqs} in lattice QCD.
It is far beyond the reach of current technology.
Here we adopt a simple toy model, which is a much simplified version of lattice effective field theory \cite{Lee:2008fa}, to demonstrate the quantum simulation described in the previous section.

Let us consider nonrelativistic fermions on a one-dimensional torus.
There are four spin-isospin species, i.e., protons and neutrons with up and down spins.
The nuclear force is approximated by contact attractive interaction.
The Hamiltonian is 
\begin{align}
\label{eqmodel}
  \hat{H} &=  Am - \frac{1}{2m} \sum_x \sum_i \hat{b}_i^\dagger(x) \partial^2 \hat{b}_i(x) - \lambda \sum_x \sum_{i<j} \hat{\rho}_i(x) \hat{\rho}_j(x)
\\
\partial^2 \hat{b}_i(x) &= \frac1{a^2} \left\{ \hat{b}_i(x+a) + \hat{b}_i(x-a) -2\hat{b}_i(x) \right\}
\\
\hat{\rho}_i(x)&= \hat{b}_i^\dagger(x) \hat{b}_i(x)
\end{align}
with the spin-isospin index $i=1,2,3,4$.
The Hamiltonian \eqref{eqmodel} preserves the fermion numbers $A_i=\sum_x \langle\hat{\rho}_i(x)\rangle$ and the total fermion number $A=\sum_i A_i$.
The energy spectrum of each fermion number sector can be exactly computed by matrix diagonalization, as shown in Fig.~\ref{figE}.
We considered the four sectors $A=2$ ($A_1=A_2=1$, $A_3=A_4=0$), $A=3$ ($A_1=A_2=A_3=1$, $A_4=0$), and $A=4$ ($A_1=A_2=A_3=A_4=1$).
As the total fermion number increases from $A=2$ to $A=4$, the energy gap between the ground state and the first excited state decreases and approximately reproduces $\Delta = O(A^{-1})$.
At the same time, the difference between the ground state energy and the non-interacting one $Am$ increases due to the attractive interaction.
The wave function gets more localized, as can be seen from the two-point correlation function $\langle \hat{\rho}_1(0) \hat{\rho}_2(x)\rangle$ in Fig.~\ref{figE}.
In the nuclear physics language, up to four nucleons can occupy the first nuclear shell and a helium $(A=4)$ is the most tightly bound.
(Note however that we need more analysis to know whether these are really bound states.)
In this model, the ground state mass is equal to the ground state energy, $M=E=\langle \Psi_A| \hat{H} |\Psi_A \rangle$, because $\langle \Psi_0| \hat{H} |\Psi_0 \rangle$ is trivially zero.

For $A\le 4$ in Fig.~\ref{figE}, the solvable Hamiltonian can be chosen as
\begin{equation}
 \hat{H}_0 = - \frac{1}{a} \sum_i \hat{\rho}_i(0) .
\end{equation}
The corresponding initial state is $|\Phi_A\rangle = \hat{\mathcal{N}}^\dagger |0\rangle = \prod_{i=1}^A \hat{b}^\dagger_i(0) |0\rangle$, where $|0\rangle$ is the empty vacuum.
The final Hamiltonian \eqref{eqmodel} is encoded by the Jordan-Wigner transformation as
\begin{align}
\begin{split}
 \hat{H} =& \ Am - \frac{1}{4ma^2} \sum_x \sum_i \Big\{ X_i(x)X_i(x+a) + Y_i(x)Y_i(x+a)- 4\hat{\rho}_i(x) \Big\} \\
&- \lambda \sum_x \sum_{i<j} \hat{\rho}_i(x) \hat{\rho}_j(x) 
\end{split}
\\
\hat{\rho}_i(x)=& \frac12 \{ 1+Z_i(x) \}.
\end{align}
Because of a periodic boundary condition, a fermionic parity is multiplied to the boundary terms, $X_i(La)=(-1)^{A_i-1} X_i(0)$ and $Y_i(La)=(-1)^{A_i-1} Y_i(0)$.
The evolution operator is discretized by the second-order Suzuki-Trotter decomposition with a small step $\delta$,
\begin{equation}
 \exp\left(-i \int_0^S ds \hat{h}(s) \right)  = \exp(-iS Am) U(S) U(S-\delta) \cdots U(\delta) 
\end{equation}
and
\begin{equation}
\begin{split}
 U(s) &= \Bigg[ \prod_x \prod_i \exp\left\{ i\frac{\delta}{2} \frac{s}{S} \frac{1}{4ma^2} Y_i(x)Y_i(x+a) \right\}
 \exp\left\{ i\frac{\delta}{2} \frac{s}{S} \frac{1}{4ma^2} X_i(x)X_i(x+a) \right\} \Bigg]
\\
&\times \left[ \prod_i \exp\left\{ -i \delta \left( 1-\frac{s}{S}\right)\frac{1}{a}\hat{\rho}_i(0)\right\} \right] \left[ \prod_x \prod_i \exp \left\{  -i\delta \frac{s}{S} \frac{1}{ma^2} \hat{\rho}_i(x) \right\} \right] \\
&\times \left[ \prod_x \prod_{i<j} \exp \left\{ i\delta \frac{s}{S} \lambda \hat{\rho}_i(x) \hat{\rho}_j(x) \right\} \right]
\\
&\times \Bigg[ \prod_x \prod_i \exp\left\{ i\frac{\delta}{2} \frac{s}{S} \frac{1}{4ma^2} X_i(x)X_i(x+a) \right\} \exp\left\{ i\frac{\delta}{2} \frac{s}{S} \frac{1}{4ma^2} Y_i(x)Y_i(x+a) \right\} \Bigg]
.
\end{split}
\end{equation}
This can be implemented by one-qubit and two-qubit gates, and so no ancilla qubit is necessary.
The number of qubits is $D_{\rm qubit}=4V=4L$.

\begin{figure}[h]
\begin{center}
 \includegraphics[width=0.45\textwidth]{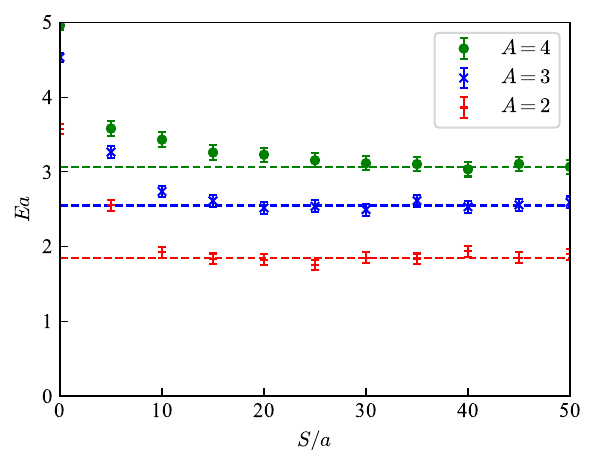}
\end{center}
\caption{
\label{figAQC}
Quantum adiabatic calculation for the ground state energy.
The broken lines are the exact values and the error bars are statistical errors.
The time step is $\delta=0.5/a$ and other conditions are the same as in Fig.~\ref{figE}.
}
\end{figure}

We numerically calculated the evolution equation \eqref{eqQAC} on a noiseless simulator.
In Fig.~\ref{figAQC}, we plot the $S$-dependence of the ground state mass.
The calculation converges at $S/a \simeq 15$ for $A=2$ and at $S/a \simeq 40$ for $A=4$.
As explained in Sec.~\ref{secqc}, the convergence is expected to scale as a function of the physical energy gap of the final Hamiltonian \eqref{eqmodel}.
From the exact results in Fig.~\ref{figE}, the ratio is $\{\Delta (A=4)/ \Delta (A=2)\}^2 \simeq (0.52/0.87)^2 \simeq 0.4$.
This is comparable with $S(A=2)/S(A=4)\simeq 0.4$.
This demonstrates the expected relation between the convergence and the physical energy gap.
The data are averaged over $10^3$ circuit executions and the standard deviation is shown as statistical error.
The statistical error is insensitive to $A$ as expected.

\section{Summary and far future applications}
\label{secF}

We have shown that the quantum simulation of lattice QCD has better dependence on the atomic number $A$ than the conventional classical simulation.
While the classical simulation suffers from crucial difficulties that
the computational cost blows up exponentially as a function of $A$,
it can be avoided by quantum simulations with the Hamiltonian formalism.
As for the dependence on the volume $V$, the quantum scaling \eqref{eqqs} can be worse than the classical scaling \eqref{eqcs}.
In practice, however, $V$ and $A$ are not independent.
The required size of $V$ will be proportional to $A$.
If the proportionality is assumed, the quantum scaling $O(A^{\frac{16}{3}})$ is always better than the classical scaling $O(e^{c_2A}A^4)$.

We have assumed fixed lattice spacing and discussed the complexity scaling of $A$ and $V$ because $A$ and $V$ are relevant parameters for nuclear many-body systems.
Of course, one eventually needs to take the continuum limit $a \to 0$.
In the classical simulation, the computational cost for large $A$ systems on a fixed physical volume 
grows at least $a^{-4}$ by the naive counting of four-dimensional lattice points,
and the actual scaling could be worse as $O(a^{-4})$ to $O(a^{-7})$
considering additional effects such as a larger autocorrelation in Monte Carlo sampling~\cite{Luscher:2010ae}.
In the quantum simulation, a similar naive counting of lattice points leads to $a^{-4}$ scaling.
However, the required value of $d$ also increases so that more ultraviolet modes are taken into account.
Consequently, the computational cost will rapidly grow \cite{Murairi:2022zdg}.
This is a crucial issue in the Hamiltonian formalism of lattice QCD and must be resolved to realize quantum advantage.

The scaling argument suggests that quantum computers are supposed to take the place of classical supercomputers at some point in the future, although the practical implementation will take a long time.
The required number of qubits is huge; e.g., $D_{\rm qubit}=D_{\rm quark}+D_{\rm gluon}\log_2 d=119{,}537{,}664$ for $V=128^3$, even if the drastic approximation $\log_2 d=11$ is adopted \cite{Alexandru:2019nsa}.
The problem of quantum error must be resolved, too.
If the implementation is achieved after a long journey, the first target would be the ground states of nuclei discussed in this paper, and then low-lying excited states.
The calculations provide not only the masses but also the state vectors.
They can be used as the state preparation for the real-time simulation of nuclear reactions, such as nucleus-nucleus scattering and nuclear fusion and fission.
Quantum computers could work as emulators of nuclear experiments.

\ack
This work was supported in part by JSPS KAKENHI Grant No.~JP19K03841, JP19K03879, JP18H05236 and JP18H05407.

\bibliographystyle{ptephy}
\bibliography{paper}

\end{document}